\documentclass[12pt]{article}
\usepackage[T1]{fontenc}
\usepackage[utf8]{inputenc}
\usepackage{lmodern}
\usepackage{rotating,amsmath,amssymb,amsfonts,amsthm,longtable}
\usepackage{epsfig,mathrsfs,natbib,color,graphics,bm,dsfont,subfigure}
\usepackage{multirow}
\def\boxit#1{\vbox{\hrule\hbox{\vrule\kern6pt \vbox{\kern6pt#1\kern5pt}
\kern6pt\vrule}\hrule}}

\newcommand{\by}{{\boldsymbol y}}
\newcommand{\bx}{{\boldsymbol x}}
\newcommand{\bz}{{\boldsymbol z}}

\newcommand{\bA}{{\boldsymbol A}}
\newcommand{\bB}{{\boldsymbol B}}

\newcommand{\bL}{{\boldsymbol L}}
\newcommand{\bI}{{\boldsymbol I}}

\newcommand{\bP}{{\boldsymbol P}}
\newcommand{\bQ}{{\boldsymbol Q}}

\newcommand{\bX}{{\boldsymbol X}}
\newcommand{\bY}{{\boldsymbol Y}}
\newcommand{\bZ}{{\boldsymbol Z}}

\newcommand{\bbeta}{{\boldsymbol \beta}}
\newcommand{\btheta}{{\boldsymbol \theta}}

\newcommand{\bepsilon}{{\boldsymbol \epsilon}}
\newcommand{\bmu}{{\boldsymbol \mu}}
\newcommand{\bpi}{{\boldsymbol \pi}}
\newcommand{\bfeta}{{\boldsymbol \eta}}
\newcommand{\bSigma}{{\boldsymbol \Sigma}}
\newcommand{\bGamma}{{\boldsymbol \Gamma}}
\newcommand{\bOmega}{{\boldsymbol \Omega}}

\def\JRSSB{{\it Journal of the Royal Statistical Society, Series B}}

\def\JRSSB{{\it Journal of the Royal Statistical Society, Series B}}

\def\JRSSB{{\it Journal of the Royal Statistical Society, Series B}}

\begin{document}
\bibliographystyle{asa}

\title{Mixture Envelope Model for Heterogeneous Genomics Data Analysis}

\author{Bochao Jia
\thanks{
 Bochao Jia is a Graduate Student, Department of Biostatistics, University of Florida, Gainesville, FL 32611,
 Email: jbc409@ufl.edu.
 }
 }

\maketitle
\renewcommand{\abstractname}{\centering\bf{Abstract}}
\begin{abstract}
\noindent
Envelope model also known as multivariate regression model was proposed to solve the multiple response regression problems. It measures the linear association between predictors and multiple responses by using the minimal reducing subspace of the covariance matrix that accommodates the mean function. However, in many real applications, data may consist many unknown confounding factors or they just come from different resources. Thus, there might be some heterogeneous dependency across the whole population and divide them into different groups. For example, there exists several subtypes across the population with breast cancer with different gene interaction mechanisms for each subtype group. In this setting, constructing a single model using all observations ignores the difference between groups while estimating multiple models for each group is infeasible due to the unknown group classification. To deal with this problem, we proposed a mixture envelope model which construct a groupwise model for heterogeneous data and simultaneously classify them into different groups by an Imputation-Conditional Consistency (ICC) algorithm. Simulation results shows that our proposed method outperforms on both classification and prediction than some existing methods. Finally, we apply our proposed method into breast cancer analysis to identify patients with inflammatory breast cancer subtype and evaluate the associations between micro-RNAs and message RNAs gene expression. 

\end{abstract}

\section{Introduction}
In cancer genomics society, microRNAs (miRNAs) play an key role in regulating gene expression at the post-transcriptional level, by binding to the $3'$ untranslated region of target messenger RNAs (mRNAs) through partial sequence homology, and causing a block of translation and/or mRNA degradation (He and Hannon, 2004). Therefore, evaluating the effect of miRNAs on mRNAs can be essential in monitoring cell differentiation, cell growth, stress response and cell death which are likely to contribute to human disease, including cancer. 

A standard model in measuring association between miRNAs and mRNAs is the multivariate linear regression given by
\begin{equation}\label{multi_reg}
\bY=\bmu +\bbeta\bX+\bepsilon
\end{equation}
where $\bY$ is an $r\times 1$ vector of multiple responses (e.g., mRNA expressions), $\bX$ is a $p\times 1$ vector of covariates (e.g., miRNA expressions), $\bmu \in \mathbb{R}^{r}$ and $\bbeta \in \mathbb{R}^{r\times p}$ are unknown intercept and regression coefficients. Moreover, the errors $\bepsilon$ follows a distribution with mean $\mathbf{0}$ and positive definite covariance matrix $\bSigma \in \mathbb{R}^{r\times r}$. In addiction to calculate the ordinary least squares estimator of $\bbeta$, Cook et al.(2010) proposed a novel envelope model framework which used the relationship among multiple responses to identify a part of them to be immaterial to $\bbeta$ and therefore reduce the variance of estimations. After the original development, further methods have been proposed to extend the scope of the envelop mode (Su and Cook, 2011,2012,2013; Cook et al., 2013; Cook and Zhang, 2015; Khare et al., 2017).

However, model (\ref{multi_reg}) cannot directly be applied to our specific problems,i.e., measuring associations between miRNAs and mRNAs since there might exists some heterogeneous dependency across the whole population and thus have different associations between miRNAs and mRNAs. For example,
miRNAs can be differentially expressed between molecular breast cancer subtypes, including luminal A, luminal B, basal-like and Her2\th. Furthermore, the regulation effect of miRNA on the expression of mRNAs also differs among subtypes (Blenkiron et al, 2007). In this settings, however, a single envelope model is not sufficient across all populations. To overcome this challenge, Park et al. (2017) proposed a groupwise envelope models for estimating heterogeneous datasets with known clusters (e.g. male vs. female). It allows for both distinct regression coefficients and distinct error structures for different groups. However, in some data sources, such as The Cancer Genome Atlas (TCGA), there might exist some unknown clusters for samples with heterogeneous models for each group.
For example, observations might have different unknown subtypes for a specific cancer and therefore, model with distinct regression parameters holds for each subtype. In this settings, groupwise envelope models can not be applied since the lack of clusters index for each observation.

To tackle this problem, some model-based cluster method can be applied which construct clusters based on the assumption that the data follows a mixture distribution. A non-exhaustive list of some works in this direction include Banfield and Raftery (1993), Biernacki et al. (1999), Fraley and Raftery (2002), Yeung et al. (2001), Medvedovic and Sivaganesan (2002), McLachlan et al. (2002), Wakefield et al. (2003) and Medvedovic et al. (2004). For example, the mixture-Gaussian model-based clustering method (MG method) is much more interesting and popular due to its simplicity in computation. This method is implemented in the R package {\emph"mclust"}, which apply the EM algorithm (Dempster et al., 1977) to estimate the model parameters, and determine the number of clusters and the covariance structure using the BIC criterion. In some real dataset, however, when the dimension of the data is high or the size of a cluster is small, it becomes difficult for parameter estimation. Moreover, the validity of the normality assumption also seldom hold. Therefore, Liang (2007) proposed a SVD-based probit transformation to overcome these problems which improves the performance of the MG method. 

Incorporated by these method, a two-stage approach can be given, i.e. apply the model-based method on $\bY$ to identify the clusters and then use groupwise envelop models for estimation. However, this two-stage approach ignores the dependency structures between predictor and responses in the clustering stage and therefore lose much accuracy for identifying clusters.

In this paper, we proposed a one-stage approach i.e, identifying cluster indices for observations and simultaneously estimate model parameters. It can be achieved by a mixture envelope model incorporated with an Imputation-Conditional Consistency (ICC) Algorithm (Liang and Jia, 2017+). Under this framework, we treat the cluster indices as the missing variables and impute them according to their posterior probabilities given the observed data. Then, we obtain a set of consistency estimates of parameters in the mixture model given the imputed values of cluster indices.

The rest of the paper is organized as follows. In section 2, we first give a brief review of envelope, groupwise envelope model and IC/ICC algorithm. Then we introduce our proposed method and its consistent properties. In section 3, we illustrate the proposed method using simulated data along with comparison with existing methods. In section 4 we apply the proposed method to a breast cancer dataset. In section 5, we conclude the paper with a brief discussion.

\section{Methods}

\subsection{A Review of Envelope and  Groupwise Envelope Models}

The original envelope model was proposed under the framework of multivariate linear regression (Cook et al., 2010). Under \ref{multi_reg}, the multiple response vector $Y$ can be partitioned into a material part and an immaterial part, where the distribution of the material part changes with the predictor $\bX$ while the immaterial part does not. In detail, let $\bL$ be an orthogonal basis of $\mathcal{S}$, where $\mathcal{S}$ is a subspace of $\mathbb{R}$ and $\bL_0$ be an orthogonal basis of $\mathcal{S}^\perp$. Then the linear combination $\bL^T\bY$ and $\bL_0^T\bY$ are called the material part and immaterial part if two conditions are satisfy: (a) $\bL^T\bY|\bX$ and $\bL_0^T\bY$ follows the same distribution and (b) $cov(\bL^T\bY,\bL_0^T\bY|\bX)=\mathbf{0}$. Let $\mathcal{B}= span(\bbeta)$, then conditions (a) and (b) are equivalent to indicate that (I) $\mathcal{B}\subseteq \mathcal{S}$ and (II) $\bSigma=\bP_{\mathcal{S}}\bSigma\bP_{\mathcal{S}}+\bQ_{\mathcal{S}}\bSigma\bQ_{\mathcal{S}}$ (Cook et al., 2010), where $\bP_{\mathcal{S}}$ denotes the projection matrix onto $span(\mathcal{S})$ and $\bQ_{\mathcal{S}}=\bI-\bP_{\mathcal{S}}$. When (I) and (II) holds, model (\ref{multi_reg}) is called the envelop model. 

Let $\mathcal{E}_{\bSigma}(\mathcal{B})$ denote the smallest reducing subspace of $\bSigma$ containing $\mathcal{B}$ which is called the $\bSigma$-envelope of $\mathcal{B}$, $u$ denote the dimension of $\mathcal{E}_{\bSigma}(\mathcal{B})$, $\bGamma \in \mathbb{R}^{r \times u}$ be an orthogonal basis of $\mathcal{E}_{\bSigma}(\mathcal{B})$, and $\bGamma_0 \in \mathbb{R}^{r \times (r-u)}$ be an orthogonal basis of $\mathcal{E}_{\bSigma}^\perp(\mathcal{B})$, the corrdinate form of envelope model can be determined as follows, 
\begin{equation}\label{envlp}
\bY=\bmu+\bGamma\bfeta\bX+\bepsilon, \quad \bSigma=\bGamma\bOmega\bGamma^T+\bGamma_0\bOmega_0\bGamma_0^T
\end{equation}
where $\bbeta=\bGamma\bfeta$, $\bfeta \in \mathbb{R}^{u \times p}$ carries the coordinates of  $\bbeta$ with respect to $\bGamma$, $\bOmega=\bGamma^T\bSigma\bGamma$ and $\bOmega_0=\bGamma_0^T\bSigma_0\bGamma_0$ carry the coordinate of  $\bSigma$ with respect to $\bGamma$ and $\bGamma_0$, respectively. Cook et al. (2010) shows that the estimates of $\bbeta$ is more efficient than or at least as efficient as the ordinary least square estimates obtained by the standard multivariate linear regression model, which can be view as the degenerated form of envelope model when $u=r$ holds.

However, when observations comes from different groups, the mixture multivariate linear regression model will be used instead of the standard ones in (\ref{multi_reg}). In detail, suppose that we observe data from $M$ different groups, for each $k=1,2,\cdots,M$, the $k$-th group has $n_k$ observations and the total sample size is $n=\sum_{k=1}^M n_k$. Then, the mixture multivariate linear regression model is defined as
\begin{equation}\label{mix_reg}
\bY_{ki}=\bmu_k+\bbeta_k\bX_{ki}+\bepsilon_{ki}
\end{equation}
where $\bY_{ki} \in \mathbb{R}^r$ is the $i$th observed response vector in the $k$th group, for $k=1,2,\cdots,M$ and $i=1,2...,n_i$. $\bmu_k \in \mathbb{R}^r$ is the mean of the $k$th group, $\bX_{ki}$ is the $i$th observed covariate vector in the $k$th group, $\bbeta_k \in \mathbb{R}^{r \times p}$ contains the regression coefficients for the $k$th group, and $\bepsilon_{ki}$ follows some distribution with mean $\mathbf{0}$ and covariance matrix $\bSigma_k$. 

Under the framework of (\ref{mix_reg}), Park et al. (2017) proposed the groupwise envelope model which extend the envelope model to multiple groups. Let $\mathcal{M}=\{\bSigma_1,\bSigma_2,\cdots,\bSigma_M\}$ denote the collection of all covariance matrices, and $\mathcal{B}=\{\bbeta_1,\bbeta_2,\cdots,\bbeta_M\}$. Then the $\mathcal{M}$-envelope of $\mathcal{B}$, denoted by $\mathcal{E}_{\mathcal{M}}(\mathcal{B})$, is the smallest subspace that reduces each matrix in $\mathcal{M}$ and contains $\mathcal{B}$. When condition (A) $span(\bbeta_k)\subseteq \mathcal{E}_{\mathcal{M}}(\mathcal{B})$ and (II) $\bSigma_k=\bP_{\mathcal{E}}\bSigma_k\bP_{\mathcal{E}}+\bQ_{\mathcal{E}}\bSigma_k\bQ_{\mathcal{E}}$ hold for $k=1,2,\cdots, M$, model (\ref{mix_reg}) is called the groupwise envelope model.

Let $\bGamma \in \mathbb{R}^{r \times u}$ be an orthogonal basis of $\mathcal{M}_{\bSigma}(\mathcal{B})$, and $\bGamma_0 \in \mathbb{R}^{r \times (r-u)}$ be its completion. the corrdinate form of the groupwise envelope model can be determined by
\begin{equation}\label{genvlp}
\bY_{ki}=\bmu_k+\bGamma\bfeta_k\bX_{ki}+\bepsilon_{ki}, \quad \bSigma_k=\bGamma\bOmega_k\bGamma^T+\bGamma_0\bOmega_0\bGamma_0^T
\end{equation}
for each $k=1,2,\cdots, M$, where $\bbeta_k=\bGamma\bfeta_k$, $\bfeta_k \in \mathbb{R}^{u \times p}$ carries the coordinates of  $\bbeta_k$ with respect to $\bGamma$, $\bOmega_k=\bGamma^T\bSigma_k\bGamma$ and $\bOmega_0=\bGamma_0^T\bSigma_0\bGamma_0$ carry the coordinate of  $\bSigma_k$ with respect to $\bGamma$ and $\bGamma_0$, respectively. 

The estimation procedure is under the assumption of normality where the normal likelihood function is applied. Theoretical properties in Park et al.(2017) demonstrate that all estimators are $sqrt{n}$-consistent estimators. More specifically, Let $\btheta=\{\bmu,\bfeta,\bOmega,\bOmega_0\}$ be a collection of parameters, where $\bmu=\{\bmu_1, \bmu_2,...,\bmu_M\}$, $\bfeta=\{\bfeta_1, \bfeta_2,...,\bfeta_M\}$, and $\bOmega=\{\bOmega_1, \bOmega_2,...,\bOmega_M\}$. When $\bGamma$ is fixed, the estimator of $\bmu_k,\bfeta_k,\bOmega_k,\bOmega_0$ can be written as explicit expressions of $\bGamma$. Let $\hat{\bSigma}_{res,k}=\frac{1}{n_k}\mathbb{Y}_{kc}^T\bQ_{\mathbb{X}_k}\mathbb{Y}_{kc}$, and $\hat{\bSigma}_{\bY}=\frac{1}{n}\sum_{k=1}^M\mathbb{Y}_{kc}^T\mathbb{Y}_{kc}$, where $\mathbb{X}_k \in \mathbb{R}^{n_k \times p}$ is the centered data matrix of $\bX$ and $\mathbb{Y}_{kc} \in \mathbb{R}^{n_k \times r}$ is the centered data matrix of $\bX$ for group $k$. Noticing that $\bGamma$ is an orthogonal basis of $\mathcal{E}_{\mathcal{M}}(\mathcal{B})$, we can obtain the estimator $\hat{\bGamma}$ by minimizing object functions in a $r \times u$ Grassmann manifold. Please refer to Park et al. (2017) for more detail. Then for $k=1,2,\cdots,M$, the estimators for all other parameters are given as follows:

\begin{itemize}
\item $\hat{\bmu}_k=\bar{\bY}_k$, where $\bar{\bY}_k=\frac{1}{n_k}\sum_{i=1}^{n_k}\bY_{ik}$;
\item $\hat{\bfeta}_k=\hat{\bGamma}^T(\mathbb{Y}_{kc}^T\mathbb{X}_k)(\mathbb{X}_k^T\mathbb{X}_k)^{-1}$;
\item $\hat{\bOmega}_k=\hat{\bGamma}^T\hat{\bSigma}_{res,k}\hat{\bGamma}$;
\item $\hat{\bOmega}_0=\hat{\bGamma}_0^T\hat{\bSigma}_{\bY}\hat{\bGamma}_0$, where $\hat{\bGamma}_0$ is the completion of $\hat{\bGamma}$;
\end{itemize}

\subsection{Imputation-Consistency Algorithm}
Let $X_{1},\dots,X_{n}$ be i.i.d. random samples from the distribution $f(x|\btheta)$. 
 Suppose $X_i=(Y_i,Z_i)$, $i=1,\ldots, n$, where $Y_i$ is observed but $Z_i$ is missing. 
 Let $\bX=(X_1,\ldots, X_n)$, $\bY=(Y_1,\ldots,Y_n)$ and $\bZ=(Z_1,\ldots, Z_n)$. 
 To estimate the parameters $\btheta$, the IC algorithm works as follows: 
 Starting with an initial guess $\btheta^{(0)}$, it iterates between 
 the imputation and consistency steps: 
 
\begin{itemize}
\item {\bf I-step}: Draw $\tilde{\bZ}$ from the conditional distribution 
  $h(\bz|\bY, \btheta_n^{(t)})$ given $\bY$ and the current estimate $\btheta_n^{(t)}$ of $\btheta$. 

\item {\bf C-step}: Based on the pseudo-complete data $\tilde{\bX}=(\bY,\tilde{\bZ})$, find 
   $\btheta_n^{(t+1)}$ which forms a consistent estimate of 
   \begin{equation} \label{Cequation}
   \btheta_*^{(t)}=\arg\max_{\btheta} E_{\btheta_n^{(t)}} \log f_{\btheta}(\tilde{\bx}),
   \end{equation} 
   where  $\tilde{\bx}=(\by,\tilde{\bz})$, $E_{\btheta_n^{(t)}} \log f_{\btheta}(\tilde{\bx})
   = \int \log(f(\tilde{\bx}|\btheta))f(\by|\btheta^*) h(\tilde{\bz}|\by,\btheta_n^{(t)}) d\by d \tilde{\bz}$, 
   $\btheta^*$ denotes the true value of the parameters, and $f(\by|\btheta^*)$ 
   denotes the marginal density function of $\by$. 
 \end{itemize}
 
 If $\btheta_n^{(t)}=\btheta^*$, 
 then $\btheta_*^{(t)}=\btheta^*$. In this case, maximizing $E_{\btheta^{*}} \log f_{\btheta}(\tilde{\bx})$ 
 is equivalent to finding a consistent estimate of $\btheta$.  
 Since a consistent estimation procedure of $\btheta$ is required for finding 
 the new estimate $\btheta^{(t+1)}$, we call this step a consistency step. 

 For high-dimensional problems, to find such a consistent estimator satisfying (\ref{Cequation}), a regularization 
 or dimension reduction-embeded parameter estimation method may be used.  
 For low-dimensional problems, the consistent estimator of $\btheta_*$ can be obtained by 
 maximizing the conditional expectation $Q(\btheta|\btheta^{(t)})$. In this sense, 
 the SEM algorithm can be viewed as a special case of the IC algorithm. 
 The IC algorithm is general. In principle, it can be applied to any problems 
 with missing data, regardless the dimension and distribution of the data.

\subsection{An Extension of the IC Algorithm}
 
 Like the EM algorithm, the IC algorithm is attractive only when the consistent estimate 
 can be easily obtained at each step. 
 We found that for many problems, similar to the ECM algorithm (Meng and Rubin, 1993), 
 the consistent estimate can be easilied obtained by a number of conditional consistency steps. 
 That is, we can partition the parameter $\btheta$ into a number of blocks and then find 
 the consistent estimator for each block conditional on the current estimates of other 
 blocks. 
 
 Suppose that $\btheta=(\btheta^{(1)}, \ldots, \btheta^{(k)})$ has been partitioned into 
 $k$ blocks. The imputation-conditional consistency (ICC) algorithm can be described as 
  follows:
 
 \begin{itemize}
 \item {\bf I-step}. Draw $\tilde{\bZ}$ from the conditional distribution
  $h(\bz|\bY, \btheta_n^{(t,1)}, \ldots, \btheta_n^{(t,k)})$ 
  given $\bY$ and the current estimate $(\btheta_n^{(t,1)},\ldots, \btheta_n^{(n,k)})$.

\item {\bf CC-step}. Based on the pseudo-complete data $\tilde{\bX}=(\bY,\tilde{\bZ})$, 
   do the following step:
   \begin{itemize}
    \item[(1)] Conditional on $(\btheta_n^{(t,2)}, \ldots, \btheta_n^{(t,k)})$, find $\btheta_n^{(t+1,1)}$ 
     which forms a consistent estimate of 
         \[
         \btheta_*^{(t,1)}=\arg\max_{\btheta^{(t,1)'}} E_{\btheta_n^{(t,1)},\ldots, \btheta_n^{(n,k)}} 
          \log f(\tilde{\bx}| \btheta_n^{(t,1)'}, \btheta_n^{(t,2)}, \ldots, \btheta_n^{(t,k)} ),
         \]

\vskip 0.5cm
    \item[(2)] Conditional on $(\btheta_n^{(t+1,1)}, \btheta_n^{(t,3)}, \ldots, \btheta_n^{(t,k)})$, find $\btheta_n^{(t+1,2)}$
     which forms a consistent estimate of
         \[
          \btheta_*^{(t,2)}=\arg\max_{\btheta^{(t,2)'}} E_{\btheta_n^{(t+1,1)}, \btheta_n^{(t,2)}, \btheta_n^{(t,3)},
           \ldots, \btheta_n^{(t,k)}} 
          \log f(\tilde{\bx}| \btheta_n^{(t+1,1)}, \btheta_n^{(t,2)'}, \btheta_n^{(t,3)}, \ldots, \btheta_n^{(t,k)} ),
         \]
\vskip 0.5cm
    \item[] $\ldots\ldots$
 \vskip 0.5cm   
    \item[(k)] Conditional on $(\btheta_n^{(t+1,1)}, \ldots, \btheta_n^{(t+1,k-1)})$, find $\btheta_n^{(t+1,k)}$
     which forms a consistent estimate of
         \[
          \btheta_*^{(t,k)}=\arg\max_{\btheta^{(t,k)'}} E_{\btheta_n^{(t+1,1)}, 
           \ldots, \btheta_n^{(t+1,k-1)}, \btheta_n^{(t,k)}}
          \log f(\tilde{\bx}| \btheta_n^{(t+1,1)}, \ldots, \btheta_n^{(t+1,k-1)}, \btheta_n^{(t,k)'} ),
         \]
         
where the expectation is with respect to the joint density function of $\tilde{\bx}=(\by,\bz)$
    and the subscript of $E$ gives the current estimte of $\btheta$. 
  \end{itemize}
 \end{itemize}
 
 It is easy to see that the ICC algorithm also forms a Markov chain. 
 The convergence of the Markov chain can be studied under the similar conditions as the IC algorithm. Your can refer Liang and Jia (2017+) for more detail.

\subsection{Foundation of Mixture Envelope Models}

In this section, we proposed the Mixture Envelope models for heterogeneous data analysis by assuming the pseudo likelihood function to be mixture normal distributed.

Let $n$ pairs of independent samples $(\bX_1,\bY_1), (\bX_2,\bY_2) \ldots (\bX_n,\bY_n)$ come from $M$ clusters which $M$ is known as a priori. Let $\btheta=\{\bpi,\bmu,\bfeta,\bOmega,\bOmega_0\}$ be a collection of parameters, where $\bpi=\{\pi_1, \pi_2,...,\pi_M\}$, $\bmu=\{\bmu_1, \bmu_2,...,\bmu_M\}$, $\bfeta=\{\bfeta_1, \bfeta_2,...,\bfeta_M\}$, and $\bOmega=\{\bOmega_1, \bOmega_2,...,\bOmega_M\}$. Then for a fixed dimension $u$, $u=0,...,r$, the normal density function for the mixture envelope model is given by
\begin{equation}\label{llf}
f(\btheta)=\prod_{i=1}^n\sum_{k=1}^M\pi_kf_k(\bX_i,\bY_i|\btheta_k),
\end{equation}
where $f_k(\bX_i,\bY_i|\btheta_k)$ is the density function for observations $(\bX_i, \bY_i)$ from the kth cluster and log form can be expressed as
\begin{equation}
\begin{split}
Log[f_k(\bX_i,\bY_i|\btheta_k)]= & -\frac{r}{2}log(2\pi)-\frac{1}{2}log|\bOmega_0|-\frac{1}{2}log|\bOmega_k|\\
&-\frac{1}{2}\left\{\bGamma^T(\bY_i-\bmu_k-\bGamma\bfeta_k\bX_i)\right\}^T\bOmega_k^{-1}\left\{\bGamma^T(\bY_i-\bmu_k-\bGamma\bfeta_k\bX_i)\right\}\\
&-\frac{1}{2}(\bY_i-\bmu_k)^T\bGamma_0\bOmega_0^{-1}\bGamma_0^T(\bY_i-\bmu_k)
\end{split}
\end{equation}
where $\btheta_k=\{(\bmu_k,\bfeta_k,\bOmega_k,\bOmega_0)\}$  is the collection of all unknown parameters for
$k=1,2,\ldots,M$. 

when $\bGamma$ is fixed, the estimators of $\bmu_k,\bfeta_k,\bOmega_k, \bOmega_0$ can be written as the expression of $\bGamma$ and the estimators of $\btheta$ can be obtained by maximizing the likelihood function
\begin{equation}\label{max}
\hat{\btheta}= \arg\max f(\btheta)
\end{equation}
However, this optimization problem can be quite complicated because the objective function is nonconvex and the solution is of rather high dimensionality. Fortunately, we show here that it can be efficiently achieved by applying our ICC algorithm. To this end, we consider the following “missing data” formulation. Let $\tau$ be a random variable indicating indicating which cluster $(\bX,\bY)$ come from such that
\begin{equation}\label{xx}
(\bX,\bY)|\tau=k \sim f_k(.|\btheta_k)
\end{equation}
and 
\begin{equation}
P(\tau=k)=\pi_k, \qquad k=1,\ldots,M.
\end{equation}

If we can observe the "complete data" $(\bX_i,\bY_i,\tau_i), i=1,...,n$, it can intuitively estimate $\btheta_k$ by maximizing the (\ref{llf}). Now that we can observe only  $(\bX_i,\bY_i)$s, we may treat $\tau_i$s as missing data and apply our ICC algorithm.

We illustrate this procedure in an iteration fashion which consists of the I-step and the CC-step in each iteration.

In the I-step, we calculate the conditional expectation of $\tau_i$ given $(\bX_i,\bY_i)$ and the current estimate of $\btheta$. Let $\btheta^{(t)}$ be the estimate of $\btheta$ at the $t$th iteration. Then according to the Baye's rule,
\begin{equation}\label{pp}
\gamma_{ik}^{(t)} = P(\tau_i=k|\bX_i,\bY_i;\btheta^{(t)})=\frac{\pi_k^{(t)}f_k(\bX_i,\bY_i|\btheta_k^{(t)})}{\sum_{l=1}^M\pi_l^{(t)}f_l(\bX_i,\bY_i|\btheta_l^{(t)})}.
\end{equation}
To impute the "missing data" $\tau_i$ at $t$th step, we consider it as an indicating variable for identifying which cluster $(\bX_i,\bY_i)$ come from, for $i=1,\ldots,n$ and draw a sample from 1 to $M$ based on a multinomial distribution with probability of ($\gamma_{i1}^{(t)}, \gamma_{i2}^{(t)}, \ldots,\gamma_{iM}^{(t)}$).

In the CC-step, we update the estimates of $\btheta=\{\bpi,\bmu,\bfeta,\bOmega,\bOmega_0\}$ by a set of consistent estimators.  
For each cluster $k=1,2,\dots, M$, $\pi_k$ can be estimated by
\begin{equation}\label{pi}
\hat{\pi}_k=n_k/n
\end{equation}
where $n_k$ is the number of samples in the $k$th cluster. For $\{\hat{\bmu}_k,\hat{\bfeta}_k,\hat{\bOmega}_k, \hat{\bOmega}_0\}$, they can be obtained by solving groupwise envelope model which are all demonstrated to be $\sqrt{n}$-consistent estimators. Therefore, our ICC algorithm can be applied here.

To sum up, we have the following Algorithm to compute $\hat{\btheta}$ as defined in equation (\ref{max}):

\vskip 0.5cm
\noindent
\textbf{Algorithm 1}
\begin{itemize}
\item[(a)]  Initialize $\btheta^{(0)}=\{\bpi^{(0)},\bmu^{(0)},\bfeta^{(0)},\bOmega^{(0)},\bOmega_0^{(0)}: k=1,\ldots,M\}$.
 
\item[(b)] For each iteration, update the estimates for each mixture parameters, respectively.

\begin{itemize}
\item[(i)] I-step: Calculate the conditional expectation from equation (\ref{pp}) and impute the indicating variable $\tau_i$ for each observation.

\item[(ii)] CC-step: Update the parameters in mixture Envelope models for each cluster by a set of consistent estimators.
\end{itemize}

\item[(c)] Do iterations in step (b) until a certain convergence criterion is met.
\end{itemize}

Thus far, we have treated the number of clusters $M$ and dimension $u$ as fixed. In real applications, we can choose them by multifold cross-validation(CV) or more practical BIC criterion. The drawback of CV is the intensive computation it requires, since it will repeatedly split, estimation and evaluation the performance for many times. Therefore, we choose BIC type of criterion to determine $(M,u)$. Following Park (2017), the total number of free parameters 
can be obtained by
\begin{equation}
N(M,u)=Mr+Mup+Mu(u+1)/2+(r-u)(r-u+1)/2+u(r-u).
\end{equation}
Then for each pair of $(M,u)$, the corresponding BIC score function is defined as
\begin{equation}\label{bic}
BIC(M,u) = -2Log[f(\hat{\btheta}(M,u))]+log(n)N(M,u)
\end{equation}
where $f(\hat{\btheta}(M,u))$ is the likelihood function given by equation (\ref{llf}). Let $(\hat{M},\hat{u})$ be the pair with smallest BIC score, and we let $\hat{\btheta}(M,u)$ as our final estimators in the mixture Envelope model.

\section{Simulation Study}
In this section, we use Monte Carlo simulations to evaluate the finite-sample performance of the mixture envelope model and compare it with standard multivariate linear regression model and two-stage model-based cluster method. We first generated the data from model with two groups ($M = 2$), which have 40 and 60\% of the observations. We set $r=10$, $p=20$, and $u=1$. The columns of matrix $(\bGamma ,\bGamma_0 )$ was generated by the eigenvectors from an $r\times r$ positive definite covariance matrix. $\bmu_1$ was a vector of 1 and $\bmu_2$ was a vector of 2, $\bfeta_1$ and $\bfeta_2$ were individually generated from a vector of independent $\chi_1^2$ and $\chi_2^2$ variates, respectively. Let $\bA \in \mathbb{R}^{(r-u)\times(r-u)}$ be a matrix of independent normal (1, 1) variates, $\bOmega_1$ and $\bOmega_2$ both be $\chi_1^2$ variates, and  $\bOmega_0=\bA\bA^T$. The predictors were generated from independent normal (1, 1) variates for all groups. We varied the total sample size from 300, 600 and 900.

To access the performance of our proposed method,
we mainly focus on the power of classification and prediction errors. To judge the quality of classification, we use the average false and negative selection rates. Let $\bm{s_k}$ denote the set of true positions of observations in cluster $k$, i.e. $\bm{s_k}=\{i:\tau_i=k\}$ where 
$\tau_i$ indicates which cluster $(\bX,\bY)$ come from and $\bm{\hat{s}_k}$ denote the set of estimated positions. Define

\begin{equation}
fsr=\frac{1}{M}\sum_{k=1}^M\frac{|\bm{\hat{s}_k}\backslash\bm{s_k}|}{|\bm{\hat{s}_k}|}, \qquad nsr=\frac{1}{M}\sum_{k=1}^M\frac{|\bm{s_k}\backslash\bm{\hat{s}_k}|}{|\bm{s_k}|}
\end{equation}
where $|\cdot|$ denotes the set cardinality and $\bA \backslash \bB$ denotes the set difference of $\bA$ and $\bB$. The smaller the values of $fsr$ and $nsr$ are, the better the performance the method is. In access the prediction quality, we adopt the $m$-fold cross validation to estimate the prediction error and the identity inner product was used to bind the elements in $(\bY-\hat{\bY})$. With different number of fold $m$, we calculate the average prediction error and its standard deviation for our proposed method. 

We also consider observations come from three clusters $M=3$. In this setting, each cluster has the same number of observations and differs from $\bmu_k$ and $\bfeta_k$, for $k=1,2,3$. We set $\bmu_1$, $\bmu_2$, $\bmu_3$ be the vector of 1,2,3, respectively and $\bfeta_1$, $\bfeta_2$, $\bfeta_3$ from a vector of independent $\chi_1^2$, $\chi_2^2$ and $\chi_3^2$ variates, respectively. In order to compare with some existing methods, we also apply standard multivariate linear regression model and two-stage approach here. For standard multivariate linear regression model (abbreviated as standard model), we can set dimension $u=r$ in the mixture envelop model which can obtain the Ordinary Least Square (OLS) estimators within each ICC iteration. For the two-stage approach, we adopt the method from Liang (2007) for identifying clusters which includes SVD decomposition of response matrix, probit transformation and model-based clustering.   

\begin{table}[!ht]
\begin{center}
\caption{Result of the our proposed method with others over 10 simulated datasets, where "M" denotes the number of clusters and "$\mbox{err}_{\bY}$" denotes the average prediction error and its standard deviation, "SVD" denotes the two-stage approach, "OLS" denotes the ICC algorithm incorporated with OLS estimators, "ICC" denotes our proposed method and "True" denotes the results obtained under true cluster.}
\label{Tab1}
\vspace{2mm}
\begin{tabular}{ccccccc} \hline
 $n$ &M & & SVD &OLS&ICC & True\\ \hline
  \multirow{6}{2cm}{\centering $300$} && $\mbox{err}_{\bY}$ & 15.086(0.453) & 17.132(1.016)& 12.535(0.252) & 12.547(0.256) \\
  &2&fsr & 0.264(0.023)& 0.391(0.026)& 0(0) & 0(0)\\ 
  && nsr & 0.263(0.024) & 0.390(0.027)& 0(0) & 0(0)\\ \cline{2-7}
    && $\mbox{err}_{\bY}$  & 14.234(0.338)  & 16.391(0.443)& 13.850(0.592)& 12.697(0.186)\\ 
  &3&fsr &0.525(0.017)& 0.600(0.008)& 0.293(0.071)& 0(0)\\ 
  && nsr & 0.545(0.013) & 0.601(0.009)& 0.299(0.071)& 0(0)\\ \hline
    \multirow{6}{2cm}{\centering $600$} && $\mbox{err}_{\bY}$ & 14.750(0.255) & 15.533(1.219)& 13.458(0.720) & 12.733(0.212)\\
  &2&fsr & 0.244(0.035) & 0.167(0.060) & 0.001(0.001)& 0(0)  \\ 
  && nsr & 0.242(0.038)& 0.163(0.060)& 0.001(0.001)& 0(0) \\ \cline{2-7}
    && $\mbox{err}_{\bY}$  & 14.476(0.516) & 15.032(0.625)& 12.792(0.470)& 12.667(0.397) \\ 
  &3&fsr & 0.480(0.022)& 0.495(0.034)& 0.074(0.051)& 0(0)\\ 
  && nsr & 0.502(0.020) & 0.494(0.036) & 0.072(0.048)& 0(0)\\ \hline  \multirow{6}{2cm}{\centering $900$} && $\mbox{err}_{\bY}$ & 14.856(0.297)& 13.862(0.734)& 12.720(0.179) & 12.720(0.180) \\
  &2&fsr & 0.243(0.052)  & 0.049(0.045)& 0.001(0.001)& 0(0)\\ 
  && nsr & 0.239(0.054)& 0.048(00.45) & 0.001(0.001)& 0(0)  \\ \cline{2-7}
    && $\mbox{err}_{\bY}$  & 15.180(0.336) & 14.645(0.303)& 13.293(0.261) & 13.252(0.263)\\ 
  &3&fsr & 0.509(0.015)& 0.389(0.050)& 0.069(0.046)& 0(0)\\ 
  && nsr & 0.516(0.016)& 0.391(0.047)& 0.067(0.044)& 0(0)\\ \hline
\end{tabular}
\end{center}
\end{table}

As showed in table \ref{Tab1}, our proposed method reaches smaller $fsr$, $nsr$ and $\mbox{err}_{\bY}$, which demonstrate to have better performance on both classification and prediction than both standard model and two stage approach.

\begin{figure}[htbp]
\centering
\subfigure[Standard deviations vs. Sample size for the elements in $\bbeta_1$ with $M=2$.]{
\label{fig1} 
\includegraphics[width=2.2in,angle=270]{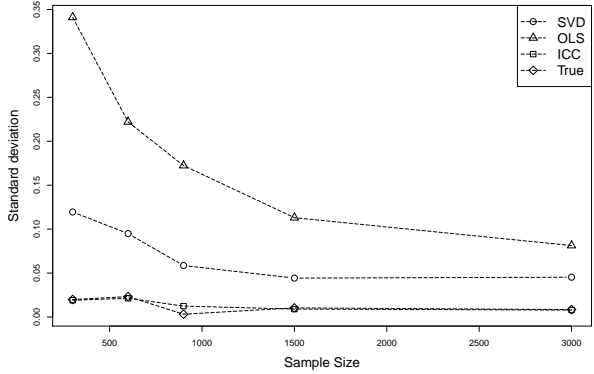}}
\hspace{0.5in}
\subfigure[Standard deviations vs. Sample size for the elements in $\bbeta_2$ with $M=2$.]{
\label{fig2}
\includegraphics[width=2.2in,angle=270]{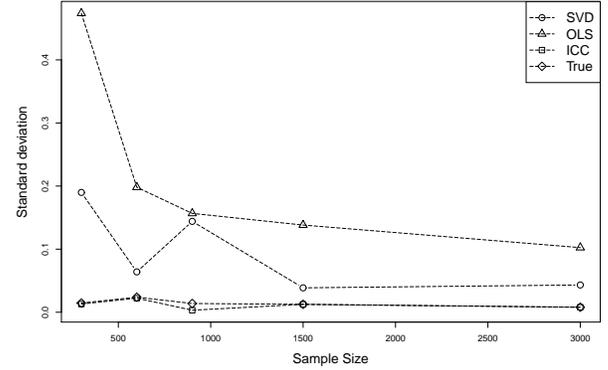}}
\\
\subfigure[Standard deviations vs. Sample size for the elements in $\bbeta_1$ with $M=3$.]{
\label{fig1} 
\includegraphics[width=2.2in,angle=270]{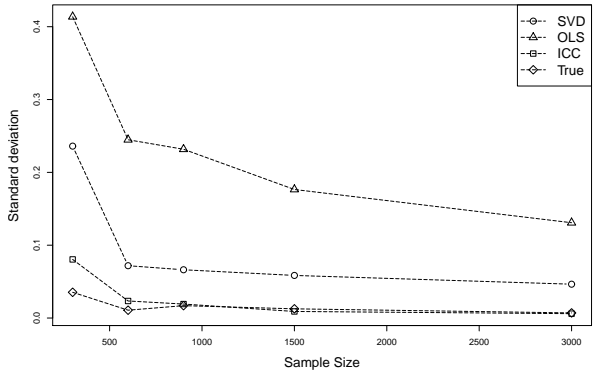}}
\hspace{0.5in}
\subfigure[Standard deviations vs. Sample size for the elements in $\bbeta_2$ with $M=3$.]{
\label{fig2}
\includegraphics[width=2.2in,angle=270]{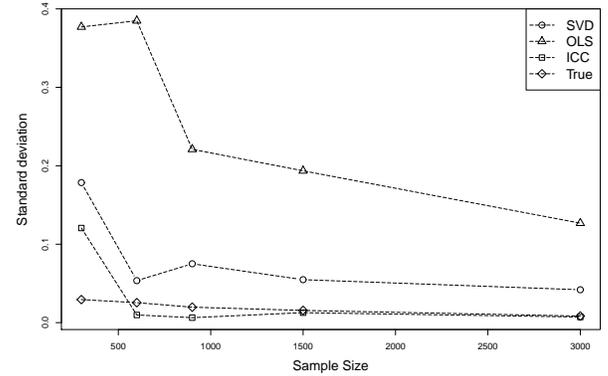}}
\\
\caption{Standard deviations vs. Sample size for the elements in $\bbeta_1$ and $\bbeta_2$. The results were obtained by two-stage method "SVD", Ordinary Least Square method "OLS", our proposed method "ICC"  and the asymptotic standard deviations under true cluster framework "True" in $M=2,3$ settings.}
\label{boot} 
\end{figure}

In order to evaluate efficiency gains for each method, standard deviation of each element in $\bbeta=(\bbeta_1,\cdots, \bbeta_M$) of model (\ref{mix_reg}) was calculated based on the 10 replications for each method at sample size of 300, 600, 900, 1500 and 3000. Since the bootstrap standard deviation is a good estimation of the sample standard deviation, we computed the bootstrap standard deviations of each element in 
$\bbeta$ based on 50 bootstrap samples.
To access the performance of each method, we calculate the the sample average of bootstrap standard deviations in each $\bbeta_k$, $k=1,2\cdots,M$ and plot the first two of them, i.e. $\bbeta_1$ and $\bbeta_2$ vs. sample sizes in Figure \ref{boot}. 

As showed in Figure \ref{boot}, our mixture envelope model achieves substantial efficiency gains over the standard multivariate linear model and SVD method. Furthermore, estimators obtained by our proposed method is consistent and their standard deviations approach to the asymptotic standard deviations under true cluster framework as sample size increases. In all panels of Figure \ref{boot}, the standard deviation of our proposed method fluctuates at a low level even with sample sizes $n=300$. This means, by using the mixture envelope model, with 300 samples, we have achieved the efficiency of taking infinity number of samples by other two methods.

Then we evaluate the performance of our proposed method for selecting the number of cluster $M$ and mixture envelope dimension $u$ where BIC criterion can be chosen.

\begin{figure}[htbp]
\centering
\subfigure[BIC score vs. number of clusters with $(M,u)=(2,1)$.]{
\label{fig1} 
\includegraphics[width=2.2in,angle=270]{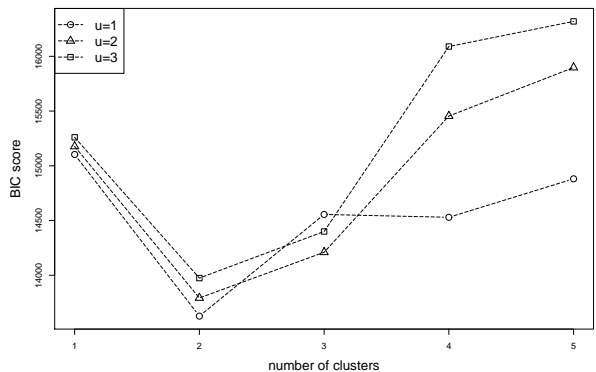}}
\hspace{0.5in}
\subfigure[BIC score vs. number of clusters with $(M,u)=(3,1)$.]{
\label{fig2}
\includegraphics[width=2.2in,angle=270]{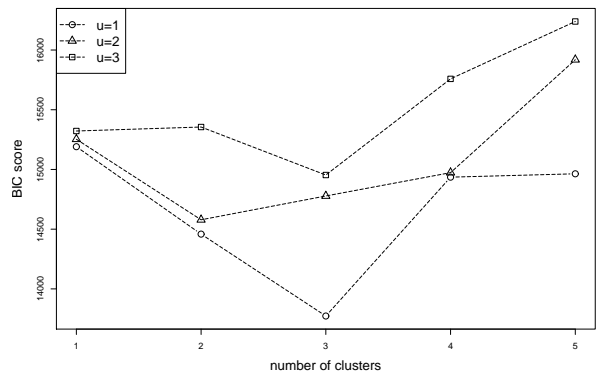}}
\\
\caption{Number of clusters $M$ and dimension $u$ selected based on BIC criterion using ICC method in the setting $n=300$.}
\label{fig_bic} 
\end{figure}

Considering $M$ to be either 2 or 3 whereas $u=1$ as the true settings.
We calculate the BIC score for each $(M,u)$ based on (\ref{bic}) and set the $(\hat{M},\hat{u})$ which corresponding to the smallest BIC score as the pair of estimators. The results are showed in Figure \ref{fig_bic}, which shows that our optimal estimators of $(\hat{M},\hat{u})$ and the true settings are exactly the same. Therefore, our proposed method can accurately discover the true number of clusters and dimension.

\section{Real Example}
In this section, we applied our mixture envelope model to the breast cancer genomics dataset and evaluate the associations between micro-RNA (miRNA) and Message RNA (mRNA) in different cancer subtypes. 
MiRNAs are a class of non-coding RNAs able to regulate mRNAs expression at the post-transcriptional level and cause a block of translation or mRNA degradation (He and Hannon, 2004). Therefore, the association between miRNA and mRNA expression has a prominent role in tumor-suppressing and defining cancer subtypes. Among all forms of breast cancer, inflammatory breast cancer (IBC) become interesting among researches since it is the deadliest form and the most aggressive type of breast cancer.

In this case study, we focus on IBC vs. Non-IBC subtypes of breast cancer which are two potential groups for all breast cancer patients. Van et al. (2010) identified 13 miRNAs which are differentially expressed between IBC vs. Non-IBC groups. Therefore, these 13 miRNAs can be viewed as the predictor variables in our mixture envelope models. As for responses, we used 14 mRNA genes which have causal effects on developing breast cancer. The gene list are available at (http://www.breastcancer.org/risk/factors/genetics)
and the dataset are available at The Cancer Genome Atlas (TCGA) data portal (http://cancergenome.nih.gov/). The dataset contains 541 samples with total 1046 normalized miRNAs and 20502 mRNA expression genes. Based on the existing knowledge, we select the subset of 13 differentially expressed miRNAs as predictors and 14 causal mRNA genes as responses.

\begin{figure}[htbp]
\centering
\includegraphics[width=4in,angle=270]{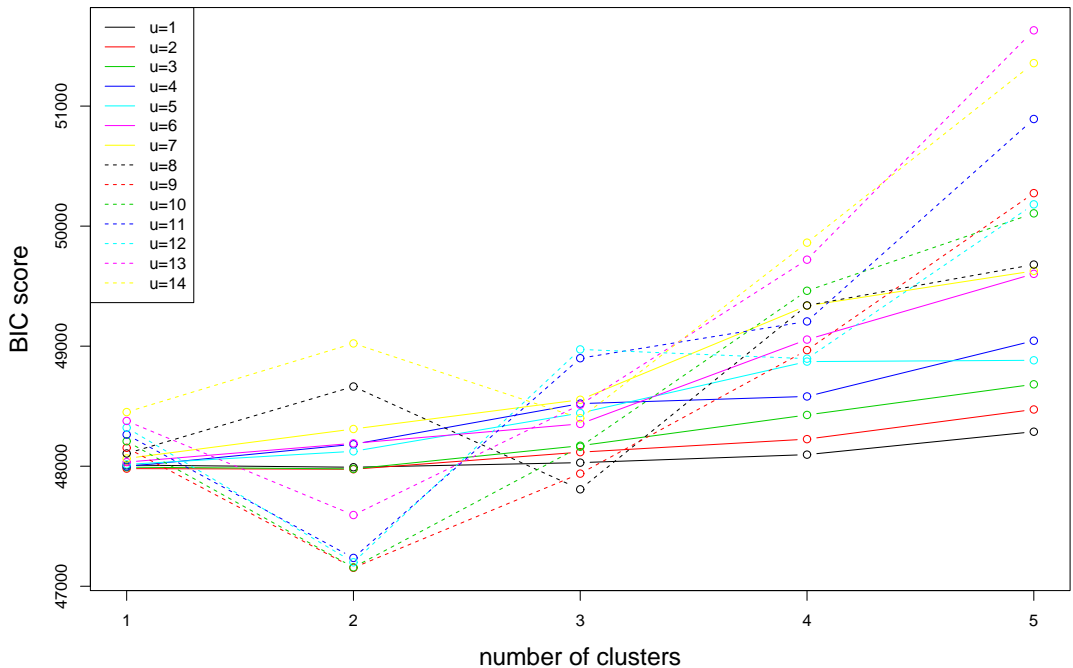}
\caption{Number of clusters $M$ and dimension $u$ selected based on BIC criterion. The optimal pair is $(\hat{M},\hat{u})=(2,9)$}
\label{bicplot}
\end{figure}

\begin{figure}[htbp]
\centering
\includegraphics[width=4in,angle=270]{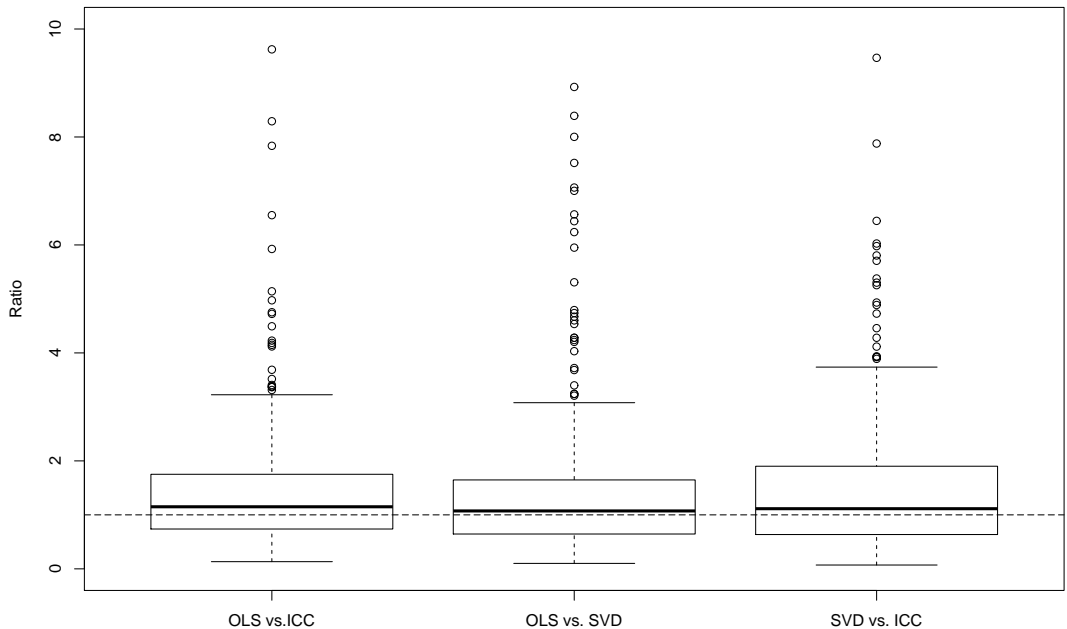}
\caption{Boxplot of ratio of the bootstrap standard deviation under standard model versus mixture envelope model, standard model versus SVD method, and SVD method versus mixture envelope model where"ICC" denotes the mixture envelope model and "OLS" denotes the standard model.}
\label{ratio_all}
\end{figure}

Then we need to determine the optimal values of number of clusters $M$ and envelope dimension $u$ by the BIC criterion. In Figure \ref{bicplot}, we plot the BIC scores vs. number of clusters under multiple values of $u=1,2,\cdots,14$. It shows that when $(M,u)=(2,9)$, BIC reaches the minimum. Therefore, the optimal number of cluster should be 2, which agrees our assumption that all patients can be divided into IBC and non-IBC groups based on the associations between these 13 differential expressed miRNAs and causal mRNA genes. Also, the envelope dimension equals to 9 which illustrate that it should obtain some efficiency gains compared with standard model. The efficiency gains can be explained by the covariance structure: $||\hat{\bOmega_1}||=2869.423$ ,$||\hat{\bOmega_2}||=7244.9$, and $||\hat{\bOmega_0}||=68.994$. This indicates that our proposed method achieve some efficiency gains in estimation. In comparison, we also applied standard model and SVD method to compute the ratio of the bootstrap standard deviation under these two methods versus the bootstrap standard deviation under mixture envelope model. Figure \ref{ratio_all} shows that our proposed method only sightly outperforms other two methods. Since $||\hat{\bOmega_0}|| \ll ||\hat{\bOmega_1}||$ and $||\hat{\bOmega_0}|| \ll ||\hat{\bOmega_2}||$, our proposed method can only have sightly efficiency gains for estimating coefficients.

In order to compare the performance on prediction, we estimate the prediction error by the average of 50 five fold cross validations with random splits and the identity inner product is used to bind the responses. The standard model has a prediction error of 103.549, the SVD method has the prediction error of 87.521, and the mixture envelop model has the prediction error of 84.459, which has the best performance on prediction. 

\begin{figure}[htbp]
\centering
\subfigure[Heatmaps of regression coefficients for mRNA vs. miRNAs in the first group.]{
\label{coef_fig1} 
\includegraphics[width=2.2in,angle=270]{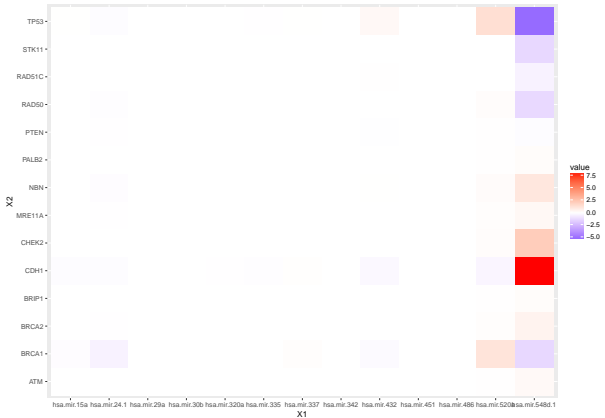}}
\hspace{0.5in}
\subfigure[Heatmaps of regression coefficients for mRNA vs. miRNAs in the second group.]{
\label{coef_fig2}
\includegraphics[width=2.2in,angle=270]{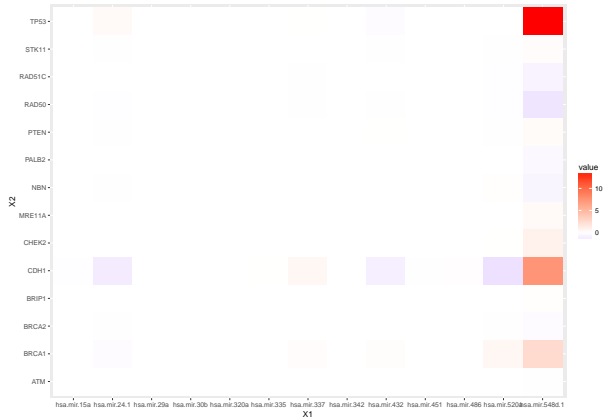}}
\\
\caption{Heatmaps of regression coefficients for mRNA vs. miRNAs. x-axis denotes the 13 differential miRNAs and y-axis denotes the 14 casual mRNA genes. }
\label{fig_coef} 
\end{figure}

\begin{figure}[htbp]
\centering
\subfigure[Plot of correlation matrix for 13 differential miRNAs for the identified Non-IBC groups.]{
\label{fig1} 
\includegraphics[width=2.2in,angle=270]{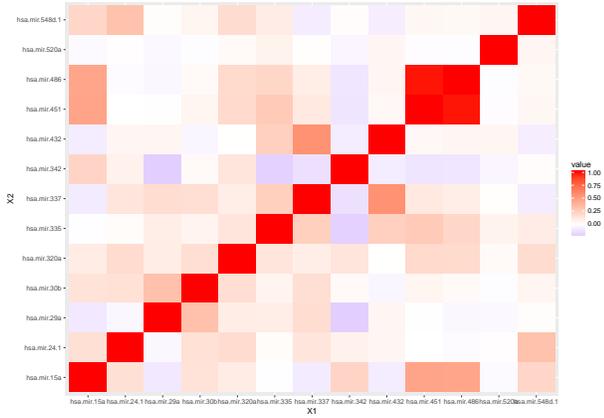}}
\hspace{0.5in}
\subfigure[Plot of correlation matrix for 13 differential miRNAs for the identified IBC groups.]{
\label{fig2}
\includegraphics[width=2.2in,angle=270]{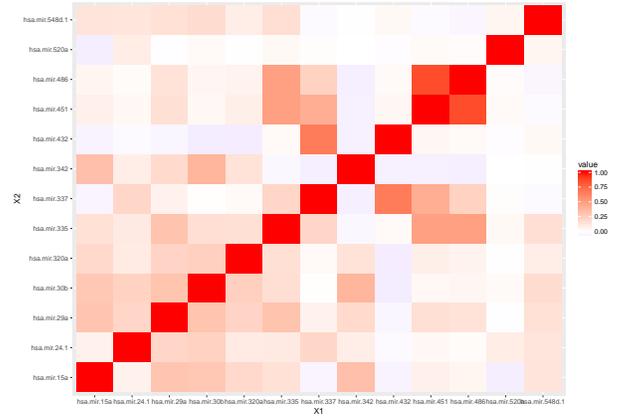}}
\\
\subfigure[Plot of correlation matrix for 14 causal gene expressions for the identified Non-IBC groups.]{
\label{fig1} 
\includegraphics[width=2.2in,angle=270]{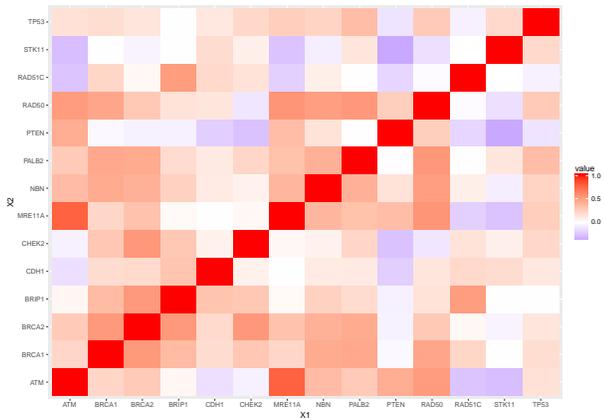}}
\hspace{0.5in}
\subfigure[Plot of correlation matrix for 14 causal gene expressions for the identified IBC groups.]{
\label{fig2}
\includegraphics[width=2.2in,angle=270]{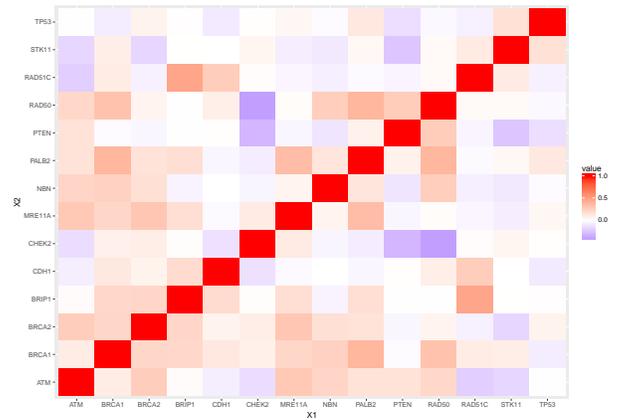}}
\\
\caption{Plot of correlation matrix for predictors (miRNAs) and responses (mRNAs).}
\label{fig_corr} 
\end{figure}

To investigate the heterogeneous dependency between two clusters, we focus on associations between miRNAs and mRNAs in two groups.
As showed in Figure \ref{fig_coef}, there exists some distinct coefficients for two groups. In Figure \ref{coef_fig1}, "mir.548d.1"  has negative coefficients on "TP53" while in Figure \ref{coef_fig2}, it has positive coefficients. Van et al. (2010) stated that the expression levels of the miR-548d.1-correlated gene sets were increased in IBC subtype. Therefore, we suggested that the second group have more likely to be with IBC subtype, since most coefficients are positive  while the first group is of Non-IBC subtype since it have some negative effects on mRNA gene expressions. Then we plot the correlation structures of predictors and responses for our identified IBC and Non-IBC groups in Figure \ref{fig_corr}, which demonstrates some heterogeneous correlation structures between two groups.

\section{Discussion}
In this paper, we proposed a mixture envelope model incorporated with Imputation Conditional Consistency Algorithm for estimating regression coefficients for heterogeneous data with unknown clusters. In addition to solve it by a two-stage method, we used a one-stage approach to identify cluster indices for observations and simultaneously estimate model parameters. In detail, assuming the data comes from multiple groups with distinct regression coefficients and heteroscedastic error structures across groups, we treated group indices as the missing variables and apply the I-step in ICC algorithm to impute them. Then the CC-step was used to estimate the $\sqrt{n}$-consistent estimator given the imputed groups indices. Our method shows great performance in both classification and predictions based on the results from simulation studies. Analysis of breast cancer shows that our proposed method have good performance for classifying patients into IBC and Non-IBC groups which is consistent with existing knowledges.

For further studies, a sparse mixture envelope model should be considered to treat high-dimensional responses. In breast cancer analysis, miRNAs might not have direct effects on the casual genes, instead, several gene subsets or some pathways of gene should be included as the responses, which might be in high-dimension. Therefore, a sparse mixture envelope model can select some casual genes from high-dimensional responses associated with miRNAs for each cancer subtype which can provide some potential biomarkers for further studies.

\section*{References} 

\begin{description}

\item[]

Banfield, J.D., Raftery, A.E., 1993. Model-based Gaussian and non-Gaussian clustering. Biometrics 49, 803–821.
\item[]
Biernacki, C., Celeux, G., Govaert, G., 1999. An improvement of the NEC criterion for assessing the number of clusters in a mixture model. Pattern Recognition Lett. 20, 267–272.

\item[]
Blenkiron, C., Goldstein, L. D., Thorne, N. P., Spiteri, I., Chin, S. F. et al. (2007). MicroRNA expression profiling of human breast cancer identifies new markers of tumor subtype. Genome biology, 8(10), R214.

\item[]
Cook, R. D., Helland, I. S., and Su, Z. (2013). Envelopes and partial least squares regression. Journal of the Royal Statistical Society: Series B (Statistical Methodology), 75(5), 851-877.

\item[]
Cook, R. D., Li, B., and Chiaromonte, F. (2010). Envelope models for parsimonious and efficient multivariate linear regression. Statistica Sinica, 927-960.

\item[]
Cook, R. D., and Zhang, X. (2015). Foundations for envelope models and methods. Journal of the American Statistical Association, 110(510), 599-611.

\item[]
Dempster, A. P., Laird, N. M., and Rubin, D. B. (1977). Maximum likelihood from incomplete data via the EM algorithm. Journal of the royal statistical society. Series B (methodological), 1-38.
\item[]

Fraley, C., Raftery, A.E., 2002. Model-based clustering, discriminant analysis, and density estimation. J. Amer. Statist. Assoc. 97, 611–631.

\item[]
He, L., and Hannon, G. J. (2004). MicroRNAs: small RNAs with a big role in gene regulation. Nature Reviews Genetics, 5(7), 522-531.

\item[]
Khare, K., Pal, S., and Su, Z. (2017). A bayesian approach for envelope models. The Annals of Statistics, 45(1), 196-222.

\item[]
Liang, F. (2007). Use of SVD-based probit transformation in clustering gene expression profiles. Computational Statistics \& Data Analysis, 51(12), 6355-6366.

\item[]
Liang,F. and Jia, B. (2017+) An Imputation-Consistency Algorithm for High-Dimensional Missing Data and Beyonds. {\it \JRSSB}.

\item[]
McLachlan, G.J., Bean, R.W., Peel, D., 2002. A mixture model-based approach to the clustering of microarray expression data. Bioinformatics 18,
413–422.

\item[]
Medvedovic, M., Sivaganesan, S., 2002. Bayesian infinite mixture model based clustering of gene expression profiles. Bioinformatics 18,
1194–1206.

\item[]
Medvedovic, M., Yeung, K.Y., Bumgarner, R.E., 2004. Bayesian mixture model based clustering of replicated microarray data. Bioinformatics 20,
1222–1232.

\item[]
Meng, X. L., and Rubin, D. B. (1993). Maximum likelihood estimation via the ECM algorithm: A general framework. Biometrika, 80(2), 267-278.

\item[]
Park, Y., Su, Z., and Zhu, H. (2017). Groupwise envelope models for imaging genetic analysis. Biometrics.

\item[]
Su, Z., and Cook, R. D. (2011). Partial envelopes for efficient estimation in multivariate linear regression. Biometrika, 98(1), 133-146.

\item[]
Su, Z., and Cook, R. D. (2012). Inner envelopes: efficient estimation in multivariate linear regression. Biometrika, 99(3), 687-702.

\item[]
Su, Z., and  Cook, R. D. (2013). Estimation of multivariate means with heteroscedastic errors using envelope models. Statistica Sinica, 213-230.

\item[]
Van der Auwera, Ilse, et al. "Integrated miRNA and mRNA expression profiling of the inflammatory breast cancer subtype." British journal of cancer 103.4 (2010): 532-541.

\item[]
Wakefield, J., Zhou, C., Self, S., 2003. Modeling gene expression over time: curve clustering with informative prior distributions. In: Bernardo, J.M.,
Bayarri, M.J., O, B.J., Dawid, A.P., Heckerman, D., Smith, A.F.M., West, M. (Eds.), Bayesian Statistics, vol. 7. Clarendon Press, Oxford.

\item[]
Yeung, K.Y., Fraley, C., Murua, A., Raftery, A.E., Ruzzo, W.L., 2001. Model-based clustering and data transformations for gene expression data.
Bioinformatics 17, 977–987.

\end{description}

\end{document}